\documentclass[conference]{IEEEtran}
\IEEEoverridecommandlockouts
\usepackage{algorithm, amsbsy,amsmath,amssymb,epsfig,bbm,mathrsfs,multirow,amsthm}
\usepackage{cite}
\usepackage{amsmath,amssymb,amsfonts}
\usepackage{graphicx}
\usepackage{textcomp}
\usepackage{subcaption} 
\usepackage{multirow}
\usepackage{bigstrut}
\usepackage{algpseudocode}
\usepackage{color}

\usepackage{array}
\newcolumntype{L}[1]{>{\raggedright\let\newline\\\arraybackslash\hspace{0pt}}m{#1}}
\newcolumntype{C}[1]{>{\centering\let\newline\\\arraybackslash\hspace{0pt}}m{#1}}
\newcolumntype{R}[1]{>{\raggedleft\let\newline\\\arraybackslash\hspace{0pt}}m{#1}}

\renewcommand{\algorithmicrequire}{\textbf{Input:}}
\renewcommand{\algorithmicensure}{\textbf{Output:}}

\def\BibTeX{{\rm B\kern-.05em{\sc i\kern-.025em b}\kern-.08em
  T\kern-.1667em\lower.7ex\hbox{E}\kern-.125emX}}
\begin{document}

\title{\huge Optimal Auction For Edge Computing
Resource Management in Mobile Blockchain Networks: A Deep Learning Approach\\
}

\author{\IEEEauthorblockN{Nguyen Cong Luong, Zehui Xiong, Ping Wang, and Dusit Niyato}
\IEEEauthorblockA{\textit{} \\
School of Computer Science and Engineering, Nanyang Technological University, Singapore 639798}
}

\maketitle
\begin{abstract}
Blockchain has recently been applied in many applications such as bitcoin, smart grid, and Internet of Things (IoT) as a public ledger of transactions. However, the use of blockchain in mobile environments is still limited because the mining process consumes too much computing and energy resources on mobile devices. Edge computing offered by the Edge Computing Service Provider can be adopted as a viable solution for offloading the mining tasks from the mobile devices, i.e., miners, in the mobile blockchain environment. However, a mechanism needs to be designed for edge resource allocation to maximize the revenue for the Edge Computing Service Provider and to ensure incentive compatibility and individual rationality is still open. In this paper, we develop an optimal auction based on deep learning for the edge resource allocation. Specifically, we construct a multi-layer neural network architecture based on an analytical solution of the optimal auction. The neural networks first perform monotone transformations of the miners' bids. Then, they calculate allocation and conditional payment rules for the miners. We use valuations of the miners as the data training to adjust parameters of the neural networks so as to optimize the loss function which is the expected, negated revenue of the Edge Computing Service Provider. We show the experimental results to confirm the benefits of using the deep learning for deriving the optimal auction for mobile blockchain with high revenue. 
\end{abstract}

%====================
\section{Introduction}
\label{Sec:intro}
%====================

% \iffalse
% which is able to record the user transactions. Blockchain outperforms centralized digital ledger approaches, in which users reach consensus of transactions with the dependency of centralized authorities, leading to high throughput and low efficiency of transaction processing. The processed transactions of users are recorded by blockchain in~\textit{blocks}, which are connected with each other as a linked list indicating the logical relations among transaction data appended to the blockchain. Accordingly, there will be no centralized authorities to store all the transaction data blocks. Instead, data blocks are recorded and shared by blockchain users over the whole blockchain network. The core aspect of ensuring the integrity of transactions recorded in blockchain is a computational process called \textit{mining}. To add new transactions to the current blockchain, a blockchain user, i.e., \textit{miner} is required to solve a computational difficult problem, i.e., proof-of-work puzzle and spread the newly revised blockchain back to the network. The rest of blockchain users will verify the validity of proof-of-work puzzle and the new blockchain. After the blockchain users reach consensus, the new transactions are successfully added as a part of the blockchain and accepted by all the users. Solving the puzzle requires high computing power and energy which restricts the use of blockchain in mobile networks. Therefore, deploying blockchain in mobile networks will become a urgent challenge in the future network.

% \fi

Blockchain has been adopted in many applications such as Bitcoin~\cite{nakamoto2008bitcoin}, smart grid power systems \cite{kang2017enabling}, and finance industry \cite{guo2016blockchain}. Recent reports predict that the annual revenue for enterprise applications of blockchain will increase from approximately \$2.5 billion worldwide in 2016 to \$19.9 billion by 2025, meaning a compound annual growth rate of 26.2\% (https://www.tractica.com/research/blockchain-for-enterprise-applications/). Different from the centralized digital ledger approaches, blockchain does not rely on centralized authorities to store transaction data. Instead, data blocks are recorded and shared by blockchain users over the whole blockchain network. Thus, the blockchain achieves high throughput and efficiency of transaction processing while maintaining data security and integrity.

%. To add new transactions to the current blockchain, a blockchain user, i.e., \textit{miner}, is required to solve a computational difficult problem, i.e., Proof-of-Work (PoW) puzzle and spread the newly revised blockchain back to the network. The processed transactions of users are recorded by blockchain in~\textit{blocks}, which are connected with each other as a linked list indicating the logical relations among transaction data appended to the blockchain. The rest of miners will verify the validity of PoW puzzle and the new blockchain. After the miners reach consensus, the new transactions are successfully added as a part of the blockchain and accepted by all the miners. 

However, deploying blockchain applications in mobile networks faces some critical challenge. This is due to the mining process, i.e., solving the \textit{Proof-of-Work (PoW)} puzzle, which requires high computing power and energy from mobile devices. To address the challenge, the edge computing paradigm is introduced into the mobile blockchain networks~\cite{youtao2017},~\cite{xiong2017} which allows the mining task of mobile users, i.e., the miners, to be offloaded to an Edge Computing Service Provider (ECSP). However, an important issue of how to efficiently allocate the limited edge computing resources to miners still remains. 

Auction becomes an appropriate solution which can guarantee that the edge computing resources are allocated to the miners which value the resources most. In a traditional auction, bidders or buyers, which are miners in the mobile blockchain context, compete for the resource units by submitting their prices, i.e., bids, to the ECSP as an auctioneer, i.e., the seller. Given the received bids, the ECSP determines the winning miners and the prices that they pay. However, traditional auctions such as the first-price auction and the second-price auction only guarantee either revenue gain for the ECSP or Incentive Compatibility (IC). The problem of designing an optimal auction in terms of maximizing the revenue for the ECSP and ensuring Dominant-Strategy IC (DSIC) as well as Individual Rationality (IR) is considerably challenging.

In recent years, the deep learning technique which is able to automatically identify relevant features has gained considerable attention. The deep learning, in principle, uses neural networks to encode any mapping from inputs to outputs~\cite{hornik1991approximation}. Especially, by using stochastic gradient descent, the deep learning succeeds in finding globally optimal solutions. In this regard, the authors in \cite{dutting2017optimal} proposed to use the deep learning for the optimal auctions. The deep learning architecture particularly fits for aforementioned setting and problem. In particular, in edge computing resource auction, the inputs to the neural networks are the miners' bids and the outputs encode the winner determination and payments of the miners. In this paper, we thus use the deep learning architecture which was proposed in \cite{dutting2017optimal} for the edge resource allocation for mobile blockchain networks. Specifically, we leverage the analytical solution from~\cite{myerson1981optimal} to construct the neural network architecture to provide precise fit to the optimal auction. The neural networks first perform monotone transformations of bidding valuations of the miners. Then, they calculate the allocation rule, i.e., winning probabilities of the miners, and the conditional payment rule to the miners. A neural network training is finally implemented to adjust parameters of the neural networks so as to optimize a loss function which is the expected, negated revenue of the ECSP. 

Simulation results demonstrate that our proposed scheme can converge quickly to the solution at which the revenue of the ECSP is significantly higher than that obtained by the traditional auction. To the best of our knowledge, this is the first paper that investigates the application of deep learning-based auction for the edge resource allocation in mobile blockchain networks. 

The rest of this paper is organized as follows. Section~\ref{Sec:related_work} reviews related work. Section~\ref{Sec:system_model} describes the system model and problem formulation. Section~\ref{Sec:DL-auction-resource-management} presents the deep learning-based optimal auction algorithm for the edge resource allocation. Section~\ref{Sec:simulation} shows the numerical performance evaluation results. Section~\ref{sec:conclusion} summarizes the paper. 

%=====================
\section{Related work}
\label{Sec:related_work}
%=====================
% proposed a game model in which the occurrence of solving the PoW puzzle is modeled as a Poisson process
There have recently been studies on the applications of game theory and pricing models for blockchain networks. As a pioneer work, the authors in \cite{kroll2013economics} modeled the mining process as a game among the miners. In the game, the strategies of the miners are to determine branches of blockchain to mine. It is then proved that if the miners behave as expected by the Bitcoin designer, there exists a Nash equilibrium in the game. The game approach is also found in \cite{houy2014bitcoin}, but the strategies of the miners are to determine the size of block to broadcast as their responses. Analytical solutions are used to prove the existence of a Nash equilibrium. However, the model has only two miners. Different from \cite{kroll2013economics} and \cite{houy2014bitcoin}, the authors in \cite{lewenberg2015bitcoin} proposed the cooperative game for the mining pool. Accordingly, the miners form a coalition to accumulate computational power and have steady reward. However, the proposed scheme only considers the internal mining, but not a dynamic environment such as the mobile blockchain network. The reason is from the fact that the mining process with high computing power demand cannot be efficiently implemented at the mobile devices. To address the issue, the authors in \cite{youtao2017} introduced an edge computing model into the mobile blockchain network. In the model, the mining process of miners, i.e., mobile users, is offloaded to an ECSP. The allocation of the edge resources to the miners is implemented using a pricing approach based on the combinatorial auction. The proposed scheme maximizes the social welfare while guaranteeing incentive compatibility or truthfulness. However, the revenue of the ECSP is not considered which is the most important factor to incentivize the ECSP to offer its edge computing services. To optimize the revenue, the authors in \cite{dutting2017optimal} proposed to use deep learning, an emerging tool for finding globally optimal solutions, for optimal auctions. However, the considered models are general auctions. 

The aforementioned approaches motivate us to investigate an optimal mechanism for the edge resource allocation in the mobile blockchain network. Specifically, we employ the deep learning architecture in \cite{dutting2017optimal} to design the optimal mechanism which guarantees the revenue maximization for the ECSP while ensuring the DSIC and IR. 

%=====================
\section{System model and problem formulation}
\label{Sec:system_model}
%=====================
This section introduces the mining process of miners and then presents the system model of the mobile blockhain network as well as the edge resource allocation problem. 

%=====================
\subsection{Blockchain Mining Process}
\label{Sec:system_model_Mining}
%=====================
\begin{figure}[h]
 \centering
\includegraphics[width=7.2cm, height = 6.5cm]{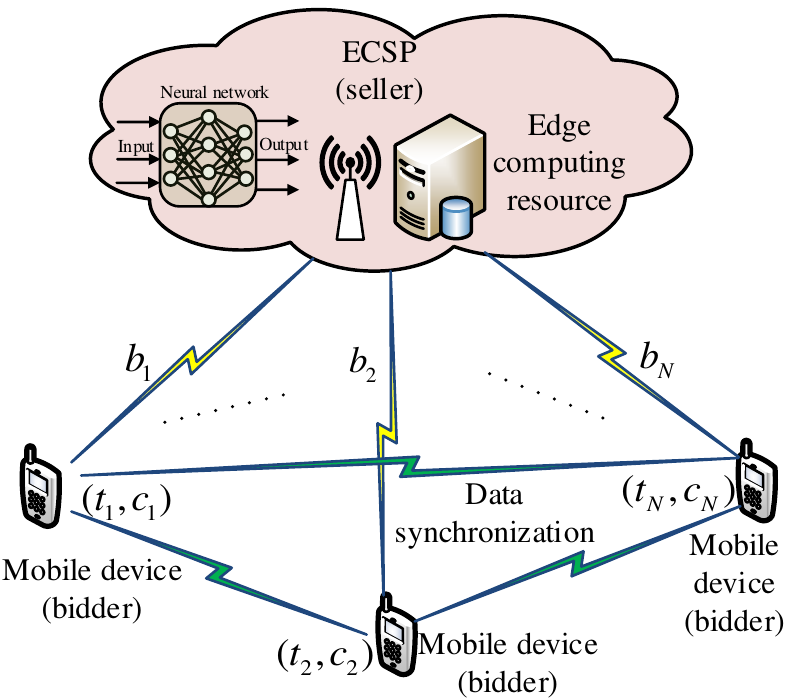}
 \caption{System model of edge computing in mobile blockchain network. ECSP stands for Edge Computing Service Provider.}
 \label{edge_computing}
\end{figure}

To create a chain of blocks, a mining process is implemented to confirm and secure transactions to be stored in a block. The mining process actually solves the proof-of-work (PoW). The PoW is a complex mathematical problem \cite{pilkington2015blockchain} in which solving the problem depends on the set of transactions to be included in the block. The mathematical problem is in a form of the hash function to combine the information about the previous block and the set of current transactions. After the problem is solved, the solution has to be propagated to reach consensus, i.e., a certain number of miners agreeing and accepting the solution. Once all these steps are done successfully, the set of transactions proposed by the miner forms a block that is appended to the current blockchain. The first miner which successfully obtains the solution of the PoW and reaches the consensus receives a mining reward. In general, solving the PoW requires high computing power, time, and energy, and thus it cannot be efficiently executed at the mobile devices. As such, we introduce the edge computing model to the mobile blockchain network for offloading the mining process from the mobile devices. 

%====================================
\subsection{Edge Computing for Blockchain Mining}
\label{sed:edge_computing_blochain_mining}
%====================================

The edge computing model is shown in Fig.~\ref{edge_computing} which consists of one ECSP and mobile users, i.e., the miners. The ECSP owns edge computing resources which are distributed across over the network to provide the mobile users computing resource services. We consider a small area of the network including $N$ mobile users and one edge computing resource unit of the ECSP. Because the edge computing resource unit is only assigned to a single mobile user, the mobile users compete on buying the unit. Note that each mobile user may already have an initial computing capacity denoted by $c_i$. If the miner obtains the edge computing resource, it will be combined with the initial computing capacity to speed up the mining process. The size of a block is denoted by $t_i$, which is the amount of transactions to be included in the block chosen by the miner. In general, when the size of the block $t_i$ is larger, the miner has more incentive to buy the edge computing resource unit to complete mining the block. This means that the miner is willing to pay a high price for the edge computing resource unit. On the contrary, when the initial computing capacity is larger, the miner has the less incentive. In this case, the miner is willing to pay a low price. Let $v_i$ denote the valuation, i.e., the private value, of miner $i$ of the edge computing resource unit. Then, $v_i$ of miner $i$ can be expressed as $v_i =\frac{t_i}{c_i}$. 

To obtain the largest revenue gain as well as to guarantee that the edge computing resource unit is allocated to the miner which values the resource unit most, the resource allocation can be modeled as a single-item auction. In the auction, the miners are bidders, i.e., the buyers, and the ECSP is the auctioneer, i.e., the seller. Miner $i$ submits price $b_i$ as a bid that the miner is willing to pay the ECSP. Based on the bid profile $\mathbf{b}=(b_1,\dots,b_N)$ from all the miners, the ECSP determines the winning miner for the edge computing unit and the corresponding price that the winner needs to pay. The ECSP can employ traditional single-item auctions such as the first-price auction and Second-Price Auction (SPA) to determine the price for the winning miner. However, none of them is an optimal auction. Specifically, the first-price auction guarantees the revenue gain for the ECSP, but cannot ensure the IC. That is, the miners have incentive to submit untruthfully their bids so as to improve their utility $u_i=v_i-b_i$, $i=1,\dots,N$. The SPA can hold the IC, but the revenue improvement for the ECSP is not guaranteed. 

Therefore, the ECSP needs to solve the problem of optimal single-item auction. More specifically, the ECSP needs to determine the winner and the corresponding payment so as to maximize the ECSP's revenue, i.e., the payment received from the winner, while guaranteeing the DSIC and IR. A mechanism is DSIC if the utility of each miner is maximized by submitting truthfully its bid regardless of the other miners' actions. The IR is to guarantee that the miners have non-negative utility for participating in the auction. The design of such an exactly optimal auction is still open. Therefore, we design the optimal single-item auction using deep learning for the resource allocation which is presented in the next section.

%=======================
\section{Optimal auction using deep learning}
\label{Sec:DL-auction-resource-management}
%=======================
% We then optimize the parameters of the neural network with the expected, negated revenue as the loss function. 
In this section, we introduce a neural network architecture for the single-item auction as presented in Section~\ref{Sec:system_model}. The neural network architecture is found in \cite{dutting2017optimal} which implements allocation and payment rules for the ECSP and guarantees that any auction mechanism learned by the network will be the optimal auction. Again, the optimal auction is in terms of maximizing revenue of the ECSP and ensuring the DSIC and IR. The learned mechanism is expected to provide precise fit to the optimal auction design, and thus we leverage the monotone transform functions, denoted as $\phi_i, i=1,\dots,N$, from \cite{myerson1981optimal} to determine the allocation and payment rules of the neural network architecture. As presented in \cite{myerson1981optimal}, input bids $b_i, i=1,\dots,N$, of miners are first transformed to $\overline{b}_i =\phi_i(b_i), i=1,\dots,N$. Then, the SPA with zero reserve price (SPA-0) is used on the transformed bids $\overline{b}_i$ to determine the allocation and conditional payment rules for the miners. Here, the reserve price refers to the lowest price which is acceptable by the ECSP for the resource unit. Let $p_i^0(\overline{b})$ and $g_i^0(\overline{b})$ denote the SPA-0 allocation rule and the SPA-0 payment rule for miner $i$, respectively. Then, we have the following theorem.

\textbf{Theorem 1} \textit{(\cite{myerson1981optimal})}\textit{.For any set of strictly monotonically increasing functions $\phi_1,\dots,\phi_N: \mathbb{R}_{\geq 0}\mapsto\mathbb{R}_{\geq 0}$, an auction which is defined by allocation rule $g_i=g_i^0 \circ \phi_i$ and conditional payment rule $p_i=\phi_i^{-1} \circ p_i^0$ is DSIC and IR.}

 Theorem 1 means that if we construct a mechanism with allocation and conditional payment rules, i.e., $g_i$ and $p_i$, then the mechanism satisfies the necessary and sufficient conditions for DSIC and IR for any choice of strictly monotone transform functions \cite{dutting2017optimal}. This is the reason that we use Theorem 1 to constrain our neural network architecture to learn the auction. As such, the auction learned by the neural network will be DSIC and IR. However, instead of specifying the precise functional form of the transform functions, the neural network learns the appropriate transform functions to minimize a loss function, i.e., the expected, negated revenue of the ECSP, which is equivalent to maximizing the expected revenue of the ECSP. The neural network architecture is shown in Fig.~\ref{neural_network}(a) with multiple layers to perform (i) the monotone transform functions $\phi_i$, (ii) the allocation rule $g_i$, and (iii) the conditional payment rule $p_i$. The algorithm for implementing these steps are given in Algorithm~\ref{alg:DL-based}, and the further details are described in the following. 
%Algorithm
%===========================
\begin{algorithm}[h]
 \caption{DL-based Auction Algorithm}
 \label{alg:DL-based}
 \begin{algorithmic}[1]
  \Require
   $N, S, J, K, \kappa, f_T(t), f_C(c),\mathbf{v}^{s}=(v_1^{s},\dots,v_N^{s}), \overline{v}_{N+1}^{s}$; 
  \Ensure Assignment probabilities ($g_1,\dots,g_N$) and conditional payments ($p_1,\dots,p_N$);
  \State Initialize: $\mathbf{w}=[w^i_{kj}] \in \mathbb{R}_{>0}^{N \times JK}$, $\boldsymbol{\beta}=[\beta^i_{kj}] \in \mathbb{R}^{N \times JK}$; \\
  \Repeat
   \State Compute $\overline{v}_i^{s}=\phi_i({v}_i^{s})=\min \limits_{1 \leq k \leq K} \max\limits_{1 \leq j \leq J} (w^i_{kj}v_i^{s} +\beta^i_{kj})$;
   \State Compute~$g_i= softmax_i(\overline{v}_1^{s},\dots,\overline{v}_{N}^{s}, \overline{v}_{N+1}^{s};\kappa)= \frac{e^{\kappa\overline{v}_i}}{\sum_{j=1}^{N+1}e^{\kappa\overline{v}_j}}$;
    \State Compute $p^0_i=ReLU \{\max\limits_{j\neq i} \overline{v}_j^{s}\}$;
    \State Compute $p_i=\phi^{-1}_i(p^0_i)=\max \limits_{1 \leq k \leq K} \min \limits_{1 \leq j \leq J} (w^i_{kj})^{-1}(p^0_i -\beta^i_{kj})$;
    \State Compute the loss function $\hat{R}(\textbf{w},\boldsymbol{\beta})= - \sum_{i=1}^N g_i^{(\textbf{w} , \boldsymbol{\beta})}(\mathbf{v}^s)  p_i^{(\textbf{w} , \boldsymbol{\beta})} (\mathbf{v}^s)$;
    \State Update $(\textbf{w},\boldsymbol{\beta})$ using the SGD solver.
  \Until The loss function $\hat{R}(\textbf{w},\boldsymbol{\beta})$ minimizes
 \end{algorithmic}
\end{algorithm}
%===========================

\begin{figure}[h]
  \begin{subfigure}[t]{0.5\linewidth}
    \centering
    \includegraphics[width=0.98\linewidth]{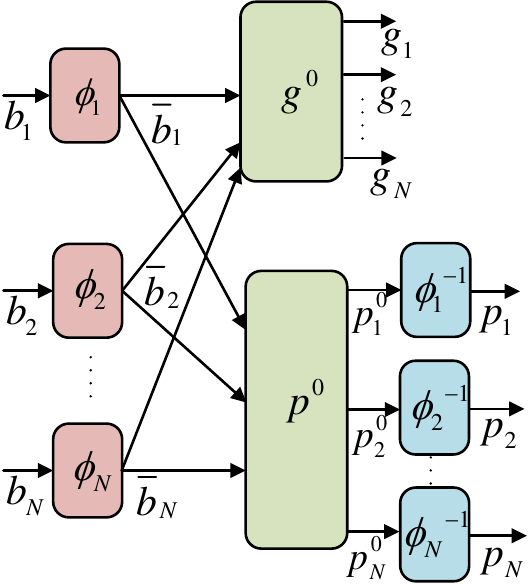}
    \caption{}
  \end{subfigure}%
 \hfill
  \begin{subfigure}[t]{0.5\linewidth}
    \centering
    \includegraphics[width=0.98\linewidth]{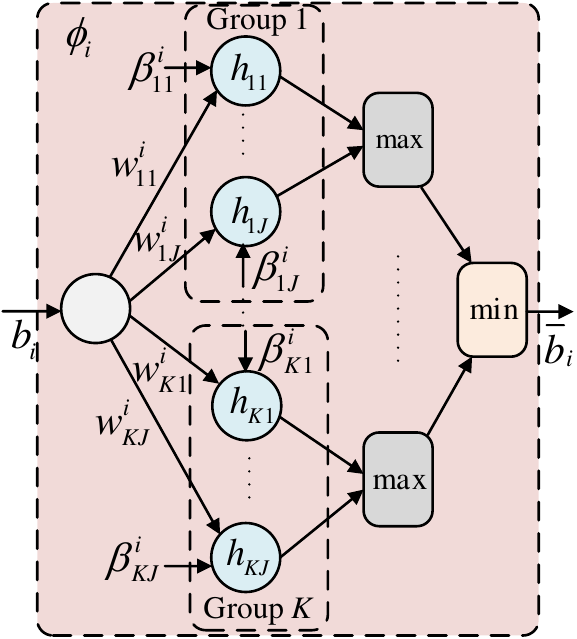}
    \large{\caption{}}
  \end{subfigure}
  \caption{Illustration of (a) neural network architecture and (b) monotone transformation $\phi_i$, where $h_{kj}(b_i)=w^i_{kj}b_i + \beta^i_{kj} $.}
  \label{neural_network}
\end{figure}

\begin{figure}[h]
  \begin{subfigure}[t]{0.5\linewidth}
    \centering
    \includegraphics[width=0.8\linewidth]{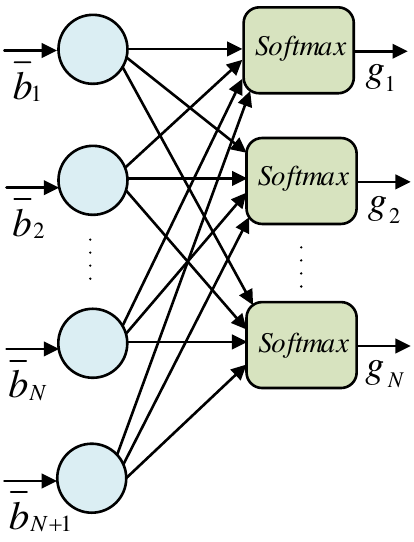}
    \caption{}
  \end{subfigure}%
 \hfill
  \begin{subfigure}[t]{0.5\linewidth}
    \centering
    \includegraphics[width=0.8\linewidth]{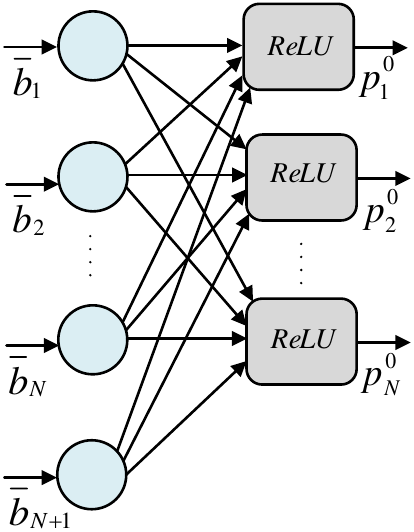}
    \large{\caption{}}
  \end{subfigure}
  \caption{Illustration of (a) allocation rule $g$ and (b) SPA-0 payment rule $p_0$.}
  \label{neural_network}
\end{figure}

%=======================
\subsection{Monotone Transform Functions}
\label{Sec:montone_transfer}
%=======================
%A graphical representation of the neural networks used for the transformation function, allocation rule, and conditional payment rule is shown in Fig.~\ref{}.
Transform function $\phi_i$ is used to map input bid $b_i$ of miner $i$ to its transformed bid $\overline{b}_i$. We model each $\phi_i$ as a two-layer feed forward network with $\min$ and $\max$ operators over linear functions as shown in Fig.~\ref{neural_network}(b). Here, we use $K$ groups of $J$ linear functions $h_{kj}(b_i)=w^i_{kj}b_i + \beta^i_{kj}$, where $k=1,\dots,K$, $j=1,\dots, J$, $w_{kj}^i \in R_{>0}$ and $\beta^i_{kj}$ are the weights and bias, respectively. Then, the transform function $\phi_i$ is defined as follows:
\begin{equation}
\phi_i(b_i) = \min_{k=1,\dots,K} \max_{j=1,\dots, J}(w_{kj}^i b_i + \beta_{kj}^i).
\label{monon_transform}
\end{equation}

In fact, the inverse transform $\phi_i^{-1}$ can be directly deduced from the parameters for the forward transform as follows:
\begin{equation}
\phi_i^{-1}(y) = \max_{k=1,\dots,K} \min_{j=1,\dots, J} (w_{kj}^i)^{-1} (y- \beta_{kj}^i).
\label{invers_monon_transform}
\end{equation}

Such a neural network is similar to a general autoencoder neural network which consists of two parts \cite{dutting2017optimal}, i.e., the encoder and the decoder. The encoder performs transformations of the input bids to a different representation through using (\ref{monon_transform}) and the decoder inverts the transform through using (\ref{invers_monon_transform}). 

%=======================
\subsection{Allocation Rule}
\label{Sec:feed-forward_winner_determination}
%=======================
The allocation rule is based on the SPA-0 allocation. Specifically, it assigns the computing resource unit of the ECSP to the miner with the highest transformed bid if this transformed bid is greater than zero, and leaves the computing resource unit unassigned otherwise. In our work, as illustrated in Fig.~\ref{neural_network}(a), the allocation rule maps the transformed bids $\mathbf{\overline{b}}=(\overline{b}_1,\dots,\overline{b}_N)$, i.e., the input, to a vector of assignment probabilities $\mathbf{g}=(g_1,\dots,g_N)$, i.e., the output. Since there is a competition among the miners, the allocation rule can be approximated by using a \textit{softmax} function on the transformed bids $\mathbf{\overline{b}}=(\overline{b}_1,\dots,\overline{b}_N)$ and an additional dummy input $\overline{b}_{N+1}=0$ as follows:
\begin{equation}
g_i(\mathbf{\overline{b}}) = softmax_i(\overline{b}_1,\dots,\overline{b}_{N+1}; \kappa) = \frac{e^{\kappa \overline{b}_i}}{\sum_{j=1}^{N+1}e^{\kappa \overline{b}_j}}, \forall i \in N,
\label{allocation_rule}
\end{equation}
where $\kappa>0$ is the parameter which determines the quality of the approximation \cite{dutting2017optimal}. The higher value of $\kappa$ increases the accuracy of the approximation. However, the allocation function may be discontinuous and less smooth which becomes harder to optimize.

%=======================
\subsection{Conditional Payment Rule}
\label{Sec:feed-forward_payment_rule}
%=======================
The conditional payment rule sets price $p_i$ to miner $i$ given that the miner is the winner. The conditional payment is implemented by two steps. The first step calculates SPA-0 payment $p_i^0$ to the miners as shown in Fig.~\ref{neural_network}(b), and the second step determines conditional payment $p_i$ by using Theorem 1. 

Specifically, the SPA-0 payment $p_i^0$ to miner $i$ is the maximum of the transformed bids from the other miners and zero. Thus it can be determined using a $ReLU$ activation unit as follows:

\begin{equation}
p_i^0(\mathbf{\overline{b}}) = ReLU(\max_{j \neq i} \overline{b}_j), \forall i \in N,
\label{payment_rule}
\end{equation}
where $ReLU(z) = \max(z, 0)$ is an activation function which ensures that the SPA-0 payment is non-negative. The conditional payment to miner $i$ is then calculated as $p_i= \phi_i^{-1} \circ p_i^0(\mathbf{\overline{b}})= \phi_i^{-1}(p_i^0(\mathbf{\overline{b}}))$, where $\phi_i^{-1}(y)$ is determined according to (\ref{invers_monon_transform}).

In summary, the allocation rule can be seen as a feed-forward network including (i) a layer of linear functions, (ii) a layer of max-min operations, and (iii) a layer of softmax activation functions. The conditional payment rule which has more layers because of using additional inverse transforms can be seen as a network consisting of (i) a layer of linear functions, (ii) a layer of max-min operations, (iii) a layer of ReLU activation functions, (iv) a layer of linear functions, and (v) a layer of min-max operations.

%=======================
\subsection{Neural Network Training}
\label{Sec:feed-forward_training_problem}
%=======================
The objective of the training is to optimize weights and bias of a neural network so as to minimize a loss function that is defined on the inputs and outputs of the network. In our work, the weights and the bias are $w_{kj}^i$ and $\beta_{kj}^i$, where $i=1,\dots,N$, $j=1,\dots,J$, and $k=1,\dots,K$, the loss function is defined as the expected, negated revenue function $\hat{R}$ of the ECSP, and the input, i.e., the training data or training set, consists of bidder valuation profiles of the miners. In particular, the bidder valuation profiles of the miners are sampled independently and identically distributed from a known distribution function. Generating the bidder valuation profiles is described as follows. 

Let $\mathbf{v}^{s}=(v_1^{s},\dots,v_N^{s})$ denote bidder valuation profile $s$ of the miners, where $s=1,\dots,S$, $S$ is the size of the training data, and $v_i^{s}$ is the valuation, i.e., the private value, of miner $i$ on the resource computing unit which is drawn from a distribution $f_V(v)$. As mentioned in Section~\ref{Sec:system_model}, $v_i^{s}$ of miner $i$ can be expressed through its block size $t_i$ and its initial computing capacity $c_i$, i.e., $v_i^{s}=t_i/c_i$. Thus, the distribution $f_V(v)$ can be determined based on the distribution of $t_i$, denoted as $f_T(t)$, and that of $c_i$, denoted by $f_C(c)$. The distributions $f_T(t)$ and $f_C(c)$ are available, e.g., based on the previous observations. Assume that variables $t$ and $c$ are independent from each other and follow uniform distributions \cite{maghsudi2016distributed}, i.e., $t \sim U[t_{min};t_{max}] $ and $c \sim U[c_{min};c_{max}]$, $c_{min} >0$. Then, $f_V(v)$ is determined as follows.

Let $z=c$, $v=t/c$, and we have $t=vz$ and $c=z$. Given the setting, the Jacobian determinant $J_D$ among $t,c,v$ and $z$ is given by 
\begin{equation}
 J_D=\begin{vmatrix} \frac{\partial t}{\partial v} & \frac{\partial t}{\partial z} \\ \frac{\partial c}{\partial v} & \frac{\partial c}{\partial z} \end{vmatrix}=\begin{vmatrix} z & v \\ 0 & 1 \end{vmatrix}= z.
 \end{equation}
 
 The Probability Density Function (PDF) for the joint distribution $(v,z)$ is given by 
\begin{align}
f_{V,Z}(v,z)=\notag &f_{T}(v,z) f_{C}(z)|J_D|\\
=&\frac{1}{(t_{max}-t_{min})(c_{max}-c_{min})} |z|.
\end{align}

The distribution of $v$, i.e., $f_{V}(v)$, is determined by
\begin{align}
f_{V}(v)=\notag &\int_{-\infty}^{+\infty} f_{V,Z}(v,z) dz\\
 = &\ \int_{c_{min}}^{c_{max}} \frac{1}{(t_{max}-t_{min})(c_{max}-c_{min})}|z|dz\\
 = \notag&\ \frac{c_{min}+c_{max}}{2(t_{max}-t_{min})},
 \label{f_V_v}
\end{align}

where $v$ is within $[t_{min}/c_{max}; t_{max}/c_{min}]$. 

Let $g_i^{(\textbf{w} , \boldsymbol{\beta})}(\mathbf{v}^s)$ and $p_i^{(\textbf{w} , \boldsymbol{\beta})} (\mathbf{v}^s)$ denote the assignment probability and conditional payment of miner $i$, respectively. $\textbf{w}$ and $\boldsymbol{\beta}$ are the matrices containing weights $w_{kj}^i$ and bias $\beta_{kj}^i$, $i=1,\dots,N$, $j=1,\dots,J$, and $k=1,\dots,K$, respectively. The objective is to find parameters $(\textbf{w}^*,\boldsymbol{\beta}^*)$ to minimize the expected, negated revenue function of the ECSP as the loss function:

\begin{equation}
\hat{R}(\textbf{w},\boldsymbol{\beta})= - \sum_{i=1}^N g_i^{(\textbf{w} , \boldsymbol{\beta})}(\mathbf{v}^s)  p_i^{(\textbf{w} , \boldsymbol{\beta})} (\mathbf{v}^s).
\label{revenue_function}
\end{equation}

We optimize the loss function $\hat{R}(\textbf{w},\boldsymbol{\beta})$ in (\ref{revenue_function}) over parameters $(\textbf{w},\boldsymbol{\beta})$ using a Stochastic Gradient Descent (SGD) solver. The implementation details are given in Section~\ref{Sec:simulation}.

%=======================
\section{Performance evaluation}
\label{Sec:simulation}
%=======================
In this section, we present experimental results to demonstrate that deep learning can be used to improve the revenue for the ECSP in the mobile blockchain network. For comparison, the proposed scheme is named Deep Learning (DL)-based auction. The SPA \cite{vickrey1961counterspeculation} is used as a baseline scheme. The DL-based auction is implemented by using the TensorFlow deep learning library. The simulation parameters are shown in Table~\ref{table:parameters}. Note that the $L_2$ regularization is used in the training step to ensure that the weight parameters are bounded. Also, the training set has 1000 valuation profiles $\mathbf{v}^s=(v_1^s,\dots,v_N^s)$, and samples $t_i$ and $c_i$ are chosen from distributions $f_T(t)$ and $f_C(c)$, respectively. 

\begin{table}[!h]
\caption{Simulation parameters} 
\centering
\begin{tabular}{ll} 
\hline\hline 
{\em Parameters} & {\em Values} \\ [0.5ex] 
\hline 
Number of miners ($N$) & 10, 15, and 20 \\ 
Learning rate & 0.0001 \\
Regularization parameter ($L_2$) & 0.01 \\
Training set size $(S)$ & 1000 valuation profiles\\
Number of groups ($K$) &5\\
Number of linear functions ($J$) &10\\
Number of iterations & 4000\\
Approximate quality $\kappa$& 1 and 2\\
Distribution of size of block $f_T(t)$ & $\sim U[0;1] $\\
Distribution of initial capacity $f_C(c)$ & $\sim U[0.2;0.5], U[0.4;0.7]$\\
\hline 
\end{tabular}
\label{table:parameters} 
\end{table}

\begin{figure}[h]
 \centering
\includegraphics[width=7cm, height = 5.6cm]{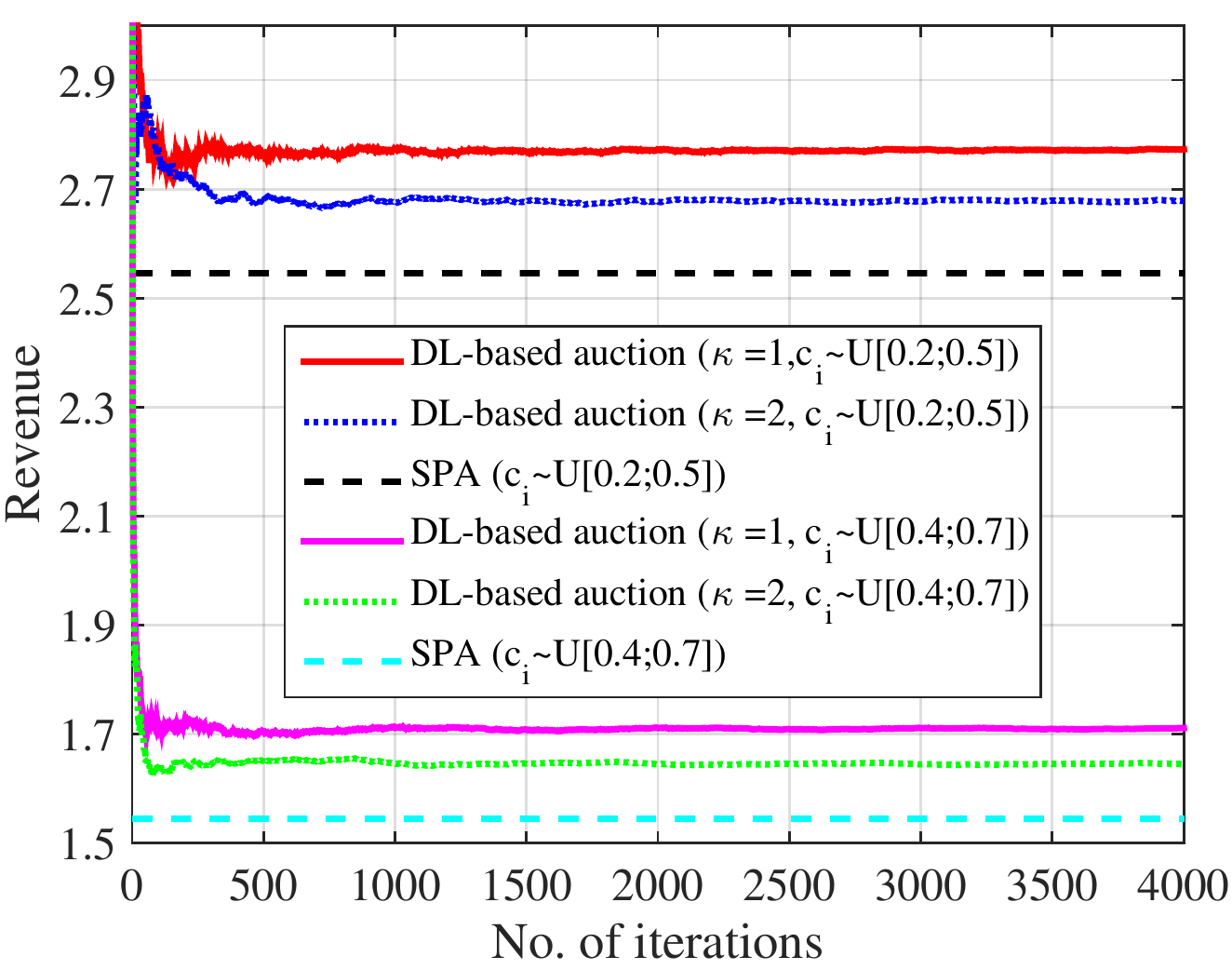}
 \caption{Test revenue for 10 miners.}
  \label{Test_revenue_10bidders}
\end{figure}

\begin{figure}[!h]
 \centering
\includegraphics[width=7cm, height = 5.6cm]{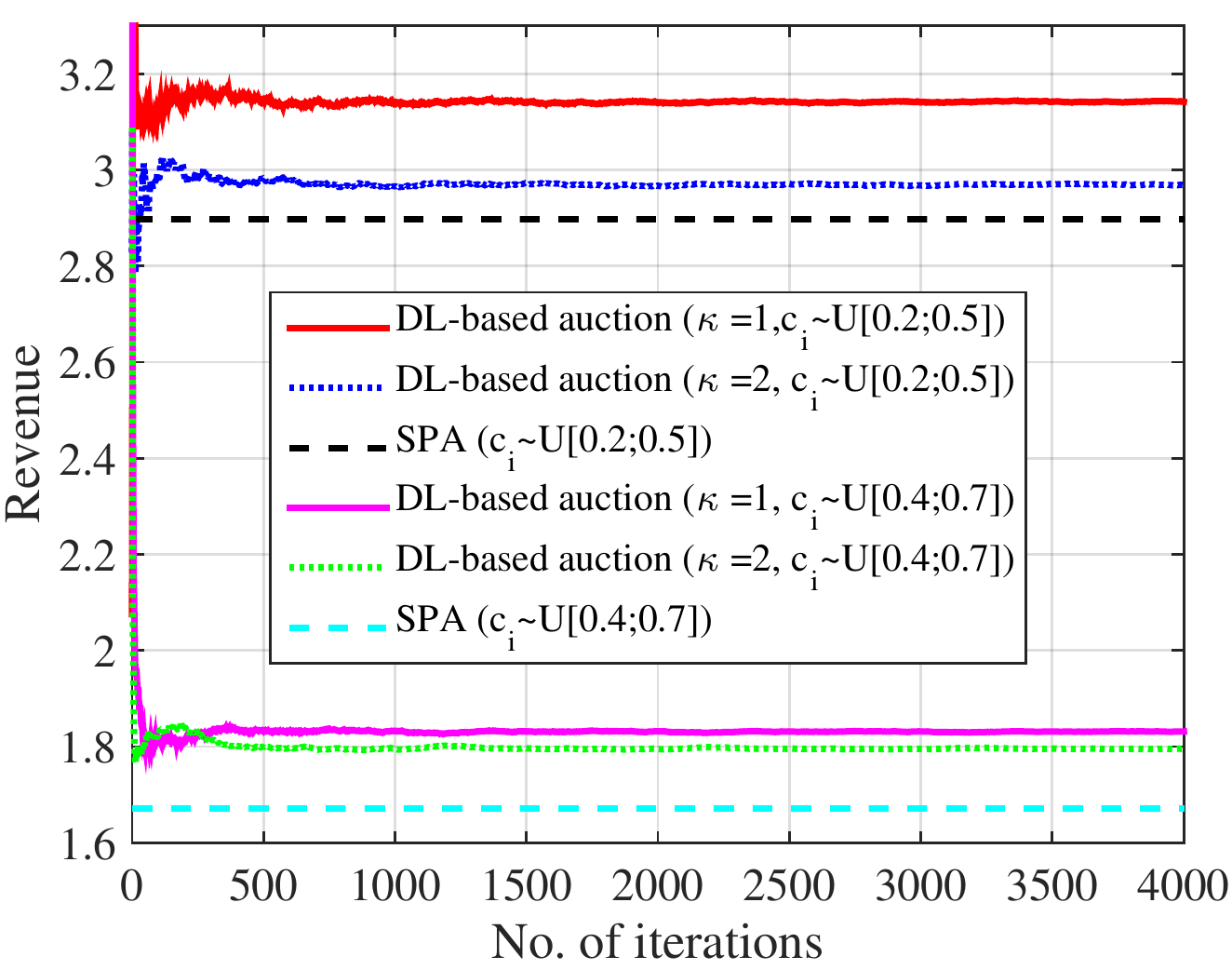}
 \caption{Test revenue for 15 miners.}
 \label{Test_revenue_15bidders}
\end{figure}

\begin{figure}[!h]
 \centering
\includegraphics[width=7cm, height = 5.6cm]{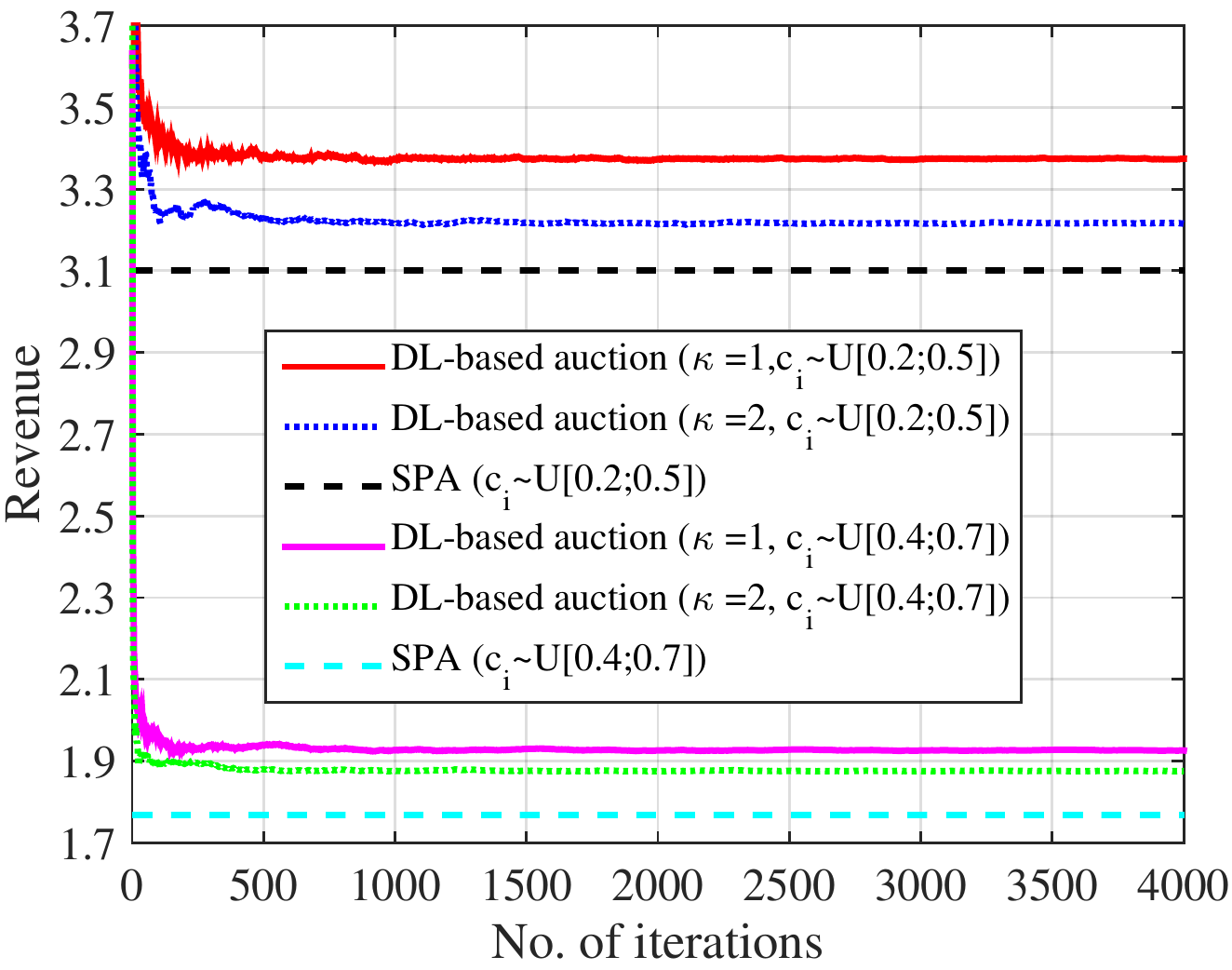}
 \caption{Test revenue for 20 miners.}
 \label{Test_revenue_20bidders}
\end{figure}

\begin{figure}[!h]
 \centering
\includegraphics[width=7cm, height = 5.6cm]{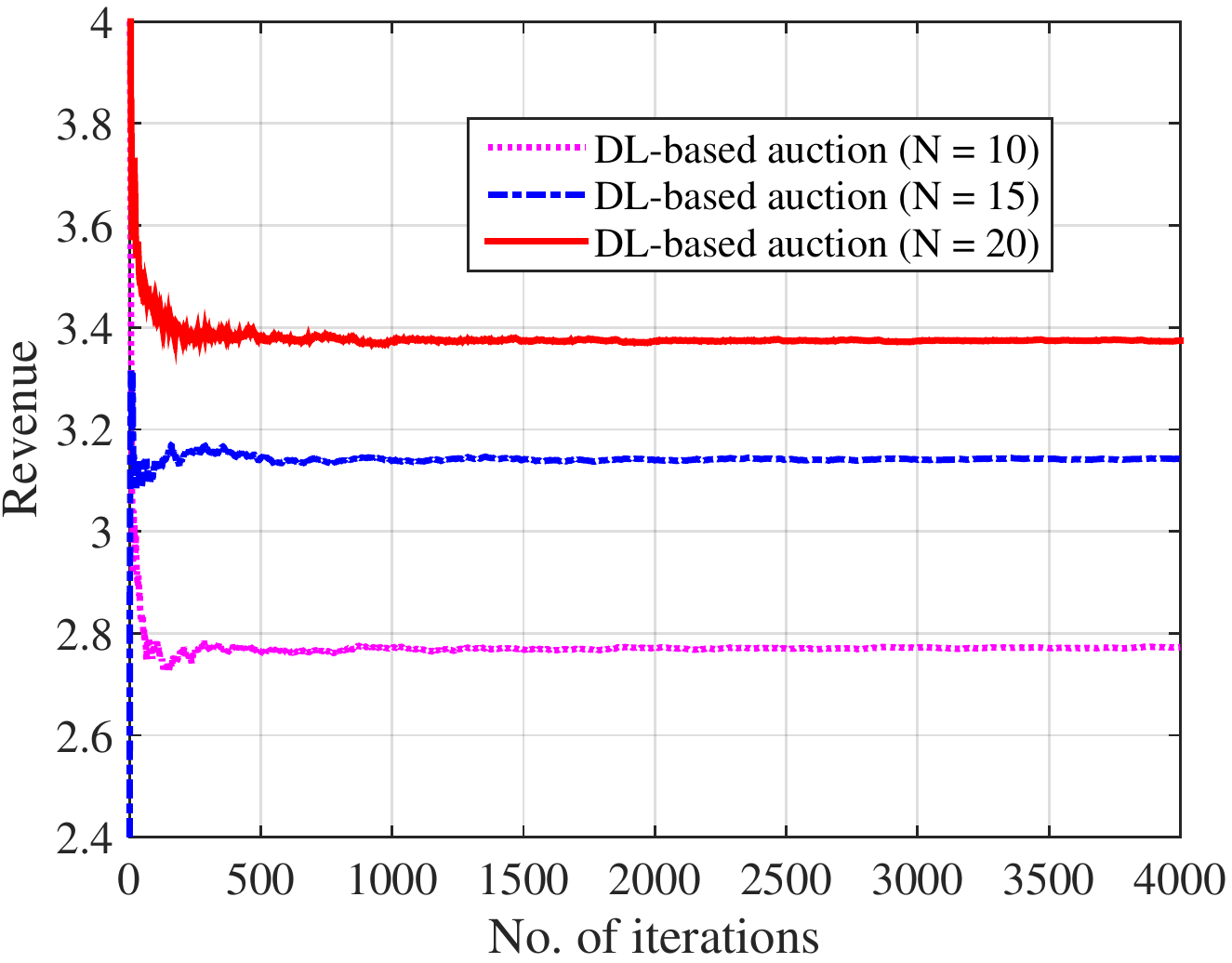}
 \caption{Test revenue for $c_i\sim U[0.2;0.5]$ and $\kappa = 1$.}
  \label{Test_revenue_10_15_20_bidders}
\end{figure}

\begin{figure}[!h]
 \centering
\includegraphics[width=7cm, height = 5.6cm]{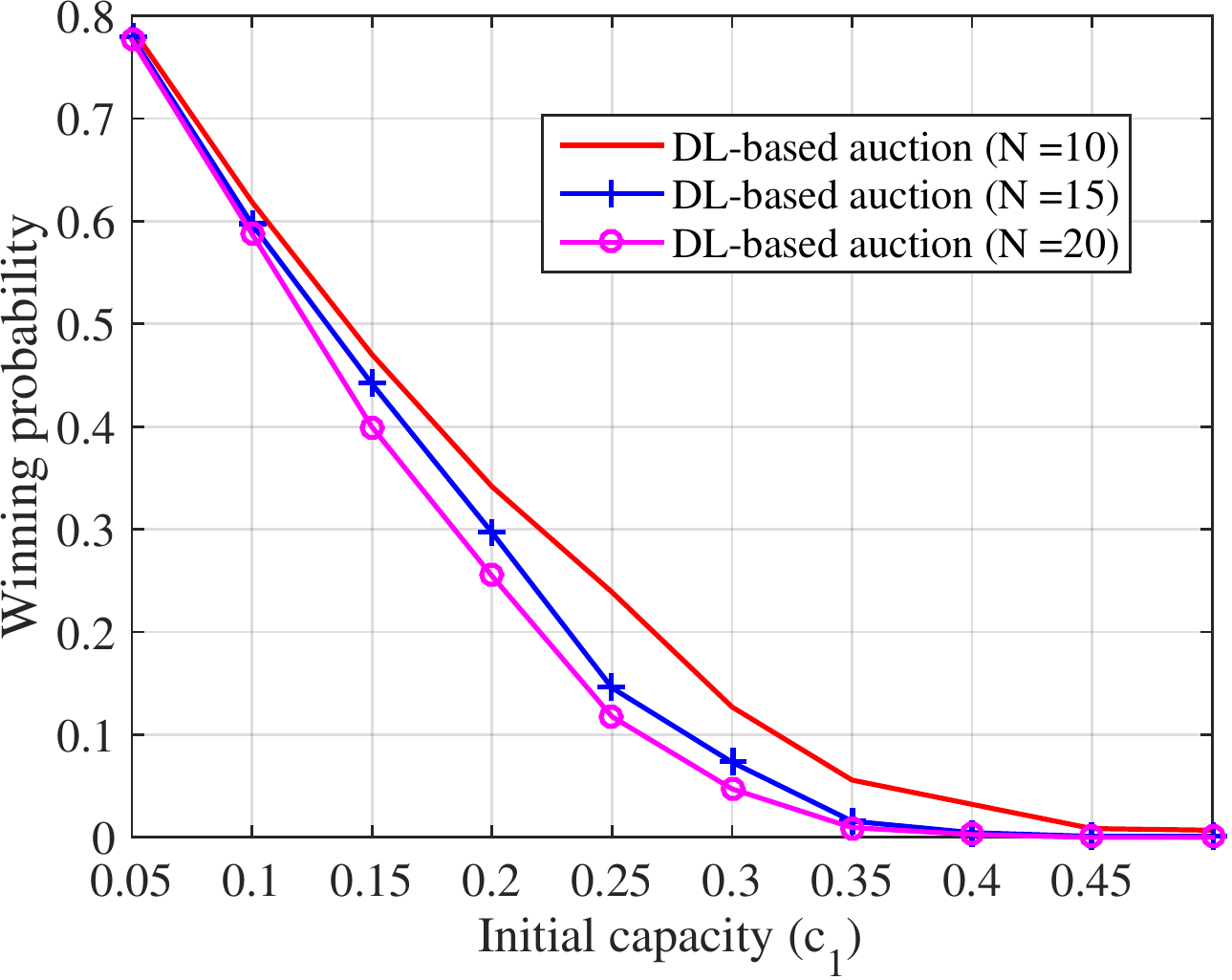}
 \caption{Winning probability of miner 1 versus initial capacity.}
  \label{winning_probability}
\end{figure}

To evaluate the performance of the DL-based auction, we consider different scenarios by varying the number of miners $N$, the distribution of initial capacity $c_i$ of the miners, and the parameter of approximate quality $\kappa$. Here, we consider the mobile blockchain network with the number of miners of 10, 15, and 20. The simulation results for the revenue versus the number of iterations are provided in Figs.~\ref{Test_revenue_10bidders},~\ref{Test_revenue_15bidders},~\ref{Test_revenue_20bidders}, and~\ref{Test_revenue_10_15_20_bidders}, and those of winning probability of the miners versus their initial capacity are shown in Fig.~\ref{winning_probability}. Note that the baseline scheme is represented by the black and cyan lines. 

It can be seen from Figs.~\ref{Test_revenue_10bidders},~\ref{Test_revenue_15bidders}, and~\ref{Test_revenue_20bidders} that the DL-based auction converges quickly to the solution which is on a par with the other schemes. Also, for a given number of miners and distribution of initial capacity, the revenue obtained by the DL-based auction is significantly higher than that obtained by the SPA. For example, for $N=15$ and $c_i\sim U[0.2;0.5]$, the revenue obtained by the SPA is 2.8966 while that obtained by the DL-based auction is 3.1460 with $\kappa =1$. The revenue improvement is clearly achieved in the other scenarios, i.e., Figs.~\ref{Test_revenue_15bidders} and~\ref{Test_revenue_20bidders}, which confirms the benefit and effectiveness of the proposed scheme. 

We next evaluate the impacts of the number of miners on the revenue of the ECSP. From Fig.~\ref{Test_revenue_10_15_20_bidders}, we find that given the distribution $c_i\sim U[0.2;0.5]$ and $\kappa=1$, the revenue of the ECSP increases with the increase of the number of miners. This is due to the fact that having more miners will intensify the competition, which potentially motivates them to pay higher service prices. As a result, the revenue of the ECSP increases. 

Then, we examine the impact of distribution ranges of initial capacity $c_i$ of miners on the revenue of the ECSP. Consider the case of 10 miners, it is observed from Fig.~\ref{Test_revenue_15bidders} that as $c_i$ is within the small range, i.e., $c_i\sim U[0.2;0.5]$, the expected revenue of the ECSP increases compared with the large range, i.e., $c_i\sim U[0.4;0.7]$. The reason is that the submitted prices, i.e., the bids, of miners are inversely proportional to the initial capacity. Therefore, given the fixed distribution of the sizes of blocks, the submitted prices are higher with the low initial capacity which in turn improves the expected revenue of the ECSP. 

Further, we consider the impact of the parameter $\kappa$ on the expected revenue of the ECSP. As mentioned in Section~\ref{Sec:DL-auction-resource-management}, $\kappa$ is introduced to the softmax function for the winner determination. In general, the large value of $\kappa$ results in a more correct decision made by the winner. However, it makes the optimization harder and more complex to solve \cite{dutting2017optimal}. This may result in reducing the expected revenue of the ECSP. Indeed, as shown in Fig.~\ref{Test_revenue_10bidders}, for $c_i\sim U[0.2;0.5]$ and $N=10$, the expected revenues of the ECSP obtained from the DL-based auction are 2.7741 and 2.6811 for $\kappa=1$ and $\kappa=2$, respectively. 

At last, it is important to consider the impact of initial capacity $c_i$ of the miners on its winning probability. Without loss of generality, we consider winning probability of miner 1 as its initial capacity $c_1$ is varied from 0.05 to 0.5 and the initial capacity of other miners is $c_i \sim U[0.2;0.5]$, $i=2,\dots,N$. With 10 miners, i.e., $N=10$, as shown in Fig.~\ref{winning_probability}, the winning probability of miner 1 decreases as its initial capacity $c_1$ increases. This is due to the fact that the miner 1's submitted price decreases. As seen, when the number of miners increases, e.g., $N=15$ and $N=20$, the winning probability of miner 1 further decreases because of more competitive miners. 

%===============
\section{Conclusions}
\label{sec:conclusion}
%===============

In this paper, we have developed an optimal auction based on deep learning for the edge resource allocation in mobile blockchain networks. Specifically, we have constructed a neural network architecture based on an analytical solution. We have designed the data training for the neural networks by using the valuations of the miners. Based on the training data, we have trained the neural networks by adjusting parameters so as to optimize the expected, negated revenue of the ECSP. As illustrated in the simulation results, the proposed scheme can quickly converge to a solution at which the revenue of the ECSP is significantly higher than that obtained by the baseline scheme. For the future work, a general scenario with multiple edge computing resource units should be considered. Also, how to construct the neural network architecture for an optimal auction without using the characterization results of the analytical solution needs to be investigated. 

\bibliographystyle{IEEEtran}
\bibliography{auction_database}
%\bibitem{Vickrey1961
%W. Vickrey, ``Counterspeculation, auctions, and competitive sealed tenders,'' \emph{The Journal of finance}, vol. 16, no. 1, pp. 8-37, 1961.

%\end{thebibliography}

\end{document}